\documentclass[showpacs,twocolumn,superscriptaddress,nofootinbib]{revtex4-1}

\usepackage{graphicx}
\usepackage{amsmath}
\usepackage{amssymb}
\usepackage{physics}
\usepackage{units}
\usepackage{color}
\usepackage{float}
\usepackage{natbib}

\newcommand{\vct}[1]{\boldsymbol{#1}}

\begin{document}

\title{Properties of a nonlinear bath: Experiments, theory, and a stochastic Prandtl-Tomlinson model}
\date{\today}

\author{Boris M\"uller}
\affiliation{Institute for Theoretical Physics, Georg-August-Universit\"at G\"ottingen, 37073 G\"ottingen, Germany}

\author{Johannes Berner}
\affiliation{Fachbereich Physik, Universit\"at Konstanz, Konstanz 78457, Germany}

\author{Clemens Bechinger}
\affiliation{Fachbereich Physik, Universit\"at Konstanz, Konstanz 78457, Germany}

\author{Matthias Kr\"uger}
\affiliation{Institute for Theoretical Physics, Georg-August-Universit\"at G\"ottingen, 37073 G\"ottingen, Germany}

\begin{abstract}
	A colloidal particle is a prominent example of a stochastic system, and, if suspended in a simple viscous liquid, very closely resembles the case of an ideal random walker. A variety of new phenomena have been observed when such colloid is suspended in a viscoelastic fluid instead, for example pronounced nonlinear responses when the viscoelastic bath is driven out of equilibrium. Here, using a micron-sized particle in a micellar solution, we investigate in detail, how these nonlinear bath properties leave their fingerprints already in equilibrium measurements, for the cases where the particle is unconfined or trapped in a harmonic potential. We find that the coefficients in an effective linear (generalized) Langevin equation show intriguing inter-dependencies, which can be shown to arise only in nonlinear baths: For example, the friction memory can depend on the external potential that acts only on the colloidal particle (as recently noted in simulations of molecular tracers in water in {\it Phys.~Rev.~X} {\bf 7}, 041065 (2017)), it can depend on the mass of the colloid, or, in an overdamped setting, on its bare diffusivity. These inter-dependencies, caused by so-called fluctuation renormalizations, are seen in an exact small time expansion of the friction memory based on microscopic starting points. Using linear response theory, they can be interpreted in terms of microrheological modes of force-controlled or velocity-controlled driving. The mentioned nonlinear markers are observed in our experiments, which are astonishingly well reproduced by a stochastic Prandtl-Tomlinson model mimicking the nonlinear viscoelastic bath. The pronounced nonlinearities seen in our experiments together with the good understanding in a simple theoretical model make this system a promising candidate for exploration of colloidal motion in nonlinear stochastic environments.
\end{abstract}

\maketitle

\section{Introduction}
Almost any matter consists of nonlinearly interacting components, giving rise to complex properties, as can be observed in prominent experiments \cite{ikeda1979multiple,faraday1831xvii,huillier1993high,morandotti1999experimental,solano2014probing}, or  computer simulations \cite{frenkel2001understanding,allen2017computer,newman1999monte,landau2014guide,tuckerman2010statistical}. This poses a serious and timely challenge of understanding such nonlinear systems, both from an experimental or phenomenological viewpoint as well as theoretically: The theoretical treatment of nonlinear systems typically requires approximations, and various schemes  have been developed \cite{davis1962introduction,khalil2002nonlinear,vidyasagar2002nonlinear,abbasbandy2007approximation,debnath2011nonlinear}. In particular nonlinear stochastic systems have proven useful in physics, chemistry, and biology, as, inter alia, in describing transition phenomena \cite{horsthemke1984noise,haken2012laser}, kinetics of phase separation \cite{mou1975kinetics,bray2002theory}, non-equilibrium thermodynamics \cite{de2013non,seifert2012stochastic}, or nonlinear fluctuational-electrodynamics \cite{soo2016fluctuational}. Regarding fluids,  a variety of formal approaches exist \cite{zwanzig1973nonlinear,klimontovich1994nonlinear,frank2005nonlinear}. 

Systems near equilibrium can generally be captured by linear (stochastic) equations, with linear coefficients renormalized by the underlying nonlinear interactions, as, e.g., exemplified by projection operator techniques \cite{mori1965continued,zwanzig1961lectures}. 

The renormalization of linear coefficients is widely known, for example regarding the linear optical response given by the permeability $\varepsilon$ \cite{Jackson}; For a typical solid, $\varepsilon$ is a function of temperature, one reason for it being  the mentioned underlying nonlinear interactions of atoms. Despite the presence of such examples, it appears that understanding of these effects in explicit experimental systems is still rare. Detecting and describing such nonlinear properties is especially important for non-equilibrium systems, as nonlinear properties also dictate the far-from-equilibrium behavior which is typically even less understood. 

In this paper, we experimentally and theoretically investigate the clean and rich system of a Brownian particle suspended in a complex, viscoelastic bath, a system which has in various forms been addressed before \cite{tseng2002micromechanical,fuchs2003schematic,solano2014probing,ciliberto2017experiments,furst2017microrheology}, and which indeed shows unexplained non-equilibrium properties \cite{gomez2015transient, berner2018oscillating}. In particular, we investigate how the nonlinear character of the bath manifests itself in equilibrium measurements, performed in presence or absence of a harmonic trapping potential, finding strong effects which can only be present in nonlinear baths. These investigations are supported by an analytical analysis within the framework of projection operator formalism: Starting from either Newtonian or overdamped dynamics, we demonstrate how microscopic interactions give rise to nontrivial dependencies (``fluctuation renormalizations'' \cite{zwanzig2001nonequilibrium}) of linear coefficients; for example the dependence of the friction kernel on the external potential, which has recently been observed for solutes in water \cite{daldrop2017external}. 
These dependencies are theoretically analyzed in several theoretical models which couple the colloidal particle to a single bath particle, finding that a stochastic Prandtl-Tomlinson model can well describe our experiments. We finally connect the equilibrium analysis to microrheology also discussing the limiting cases of weak and strong external confinement \cite{squires2005simple}.    

\section{Experiment: Non-Gaussian displacements}\label{sec:non_gaussian_displacements}
\begin{figure}[b]
\includegraphics[width=.6\linewidth]{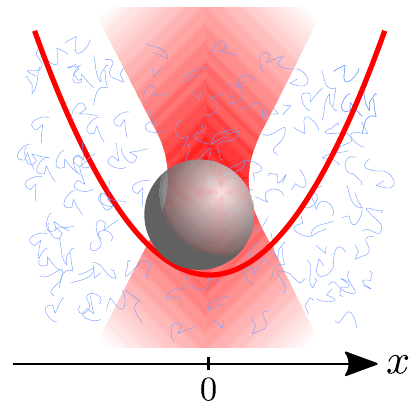}
\caption{\label{fig:schematic_exp}Experimental setup of a colloidal particle in a micellar solution subjected to a harmonic confinement potential. The micrometer-sized particle performs a typical random walk in the limited configurational space in thermal equilibrium.}
\end{figure}

The experimental analysis is performed in an equimolar solution of surfactant, cetylpyridinium chloride monohydrate (CPyCl) and sodium salicylate (NaSal) in deionised water at a concentration of $\unit[7]{mM}$ and at room temperature, $T=\unit[298\pm 0.2]{K}$. After overnight mixing, worm-like micelles form and deform dynamically in such solvents \cite{cates1990statics}. They build a highly dynamical entangled viscoelastic network which exhibits a comparatively large structural relaxation time of $\tau_s=\unit[2.5\pm 0.2]{s}$ determined by a recoil experiment \cite{gomez2015transient} and macrorheological measurements, thereby giving rise to highly non-Newtonian properties \cite{gomez2015transient, berner2018oscillating}. The length of wormlike micelles is typically found between $\unit[100]{}$ and $\unit[1000]{nm}$ \cite{walker2001rheology}, and the characteristic mesh size is on the order of \unit[30]{nm} \cite{buchanan2005high}.
We examine the thermal equilibrium fluctuations of a single mesoscopic silica particle of diameter $2R=\unit[2.73]{\mu m}$. While the particle naturally lives in a three dimensional surrounding, far from any boundaries, we concentrate on its $x$-component. The particle is trapped by a highly focused laser beam, which creates a static parabolic potential $V_\mathrm{ext}=\frac{1}{2}\kappa x^2$, with $x$ the spatial coordinate relative to the potential minimum, see Fig.~\ref{fig:schematic_exp}. The focal plane is adjusted to the middle of the sample cell, so that the trap position is more than $\unit[40]{\mu m}$ away from any walls and hydrodynamic interactions with walls can be ruled out. $x(t)$ is recorded at rates of at least 100 fps.

\begin{figure}[t]
\includegraphics[width=\linewidth]{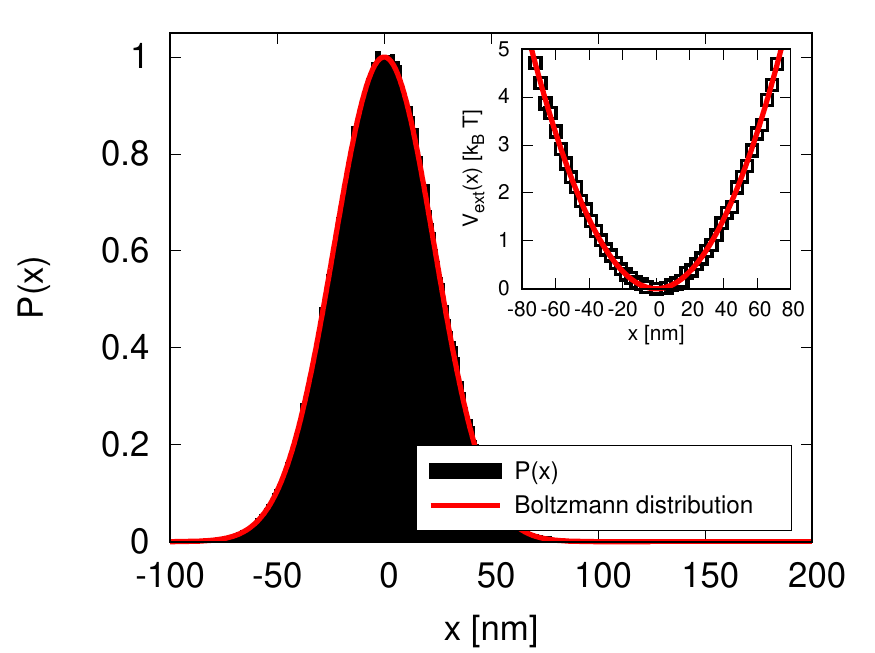}
\caption{\label{fig:x_fluctuations}Main graph: Probability distribution $P(x)$ for a harmonically trapped particle, which is well described by the Boltzmann distribution shown as a red line.
Inset: Measured trap potential (symbols) and a parabolic fit (solid line), from which the trap stiffness $\kappa$ is extracted.} 
\end{figure}

As mentioned, this system displays highly nonlinear properties when driven out of equilibrium  \cite{gomez2015transient, berner2018oscillating}, which triggers the question whether and how these nonlinear properties can already be detected in equilibrium, where the particle positions follow the Boltzmann distribution as shown in Fig.~\ref{fig:x_fluctuations}, $P(x)\propto e^{-V(x)/k_BT}$, with Boltzmann constant $k_B$ ($P(x)$ thus allows determination of the value of $\kappa$). In order to address this, we start with the case $\kappa=0$, and investigating the particle's free diffusion (again regarding the $x$ coordinate). A well-known measure of nonlinearity is then given by the incoherent dynamic structure factor \cite{dhont1996introduction}
\begin{equation}
	S_s(k,t)\equiv \langle\exp(ik(x(0)-x(t)))\rangle\equiv \exp(-D(k,t)k^2t)\,.\label{eq:Sq}
\end{equation}
The right-hand side of Eq.~\eqref{eq:Sq} defines the diffusion coefficient $D(k,t)$, which depends in general on wavevector $k$ and time $t$ (and has recently attracted a lot of interest \cite{burada2009diffusion,ernst2013measuring,dechant2019estimating}). As evident from Eq.~\eqref{eq:Sq}, a Gaussian process, as found in purely linear systems, shows no $k$-dependence in $D$, so that such $k$-dependence is a direct indicator for a non-Gaussian and nonlinear process.

\begin{figure}
\includegraphics[width=\linewidth]{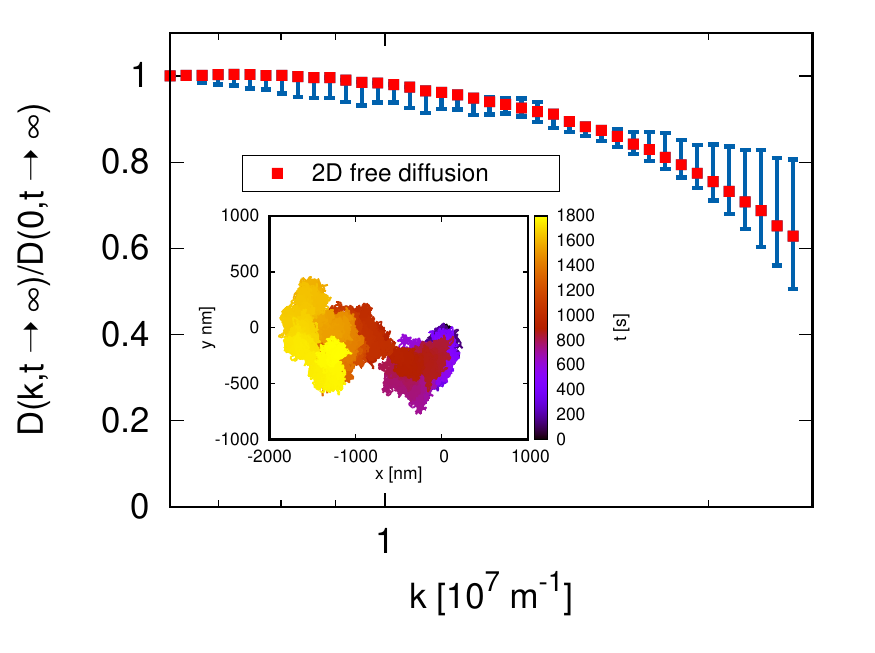}
\caption{\label{fig:diff_coeff_long_time_exp} Main graph: Normalized wavenumber-dependent long-time diffusion coefficient $D(k,t\to\infty)$ extracted from a long-time 2D free diffusion measurement of a tracer particle in a micellar bath. Error bars show the statistical error estimated from partitioning the measured trajectory into two pieces. Inset: Recorded experimental trajectory $(x(t),y(t))$ over a time period of $\unit[1800]{s}$.}
\end{figure}

Fig.~\ref{fig:diff_coeff_long_time_exp} shows the long time limit of $D(k,t)$, obtained as 
\begin{equation}\label{eq:diffcoeff_interacting}
	\lim_{t\to\infty}D(k,t)=-\lim_{t\to \infty}\frac{1}{k^2t}\log\left(\langle\exp(ik(x(0)-x(t)))\rangle\right)\,,
\end{equation}
and normalized on $D(0,t\to\infty)=\unit[204.8]{nm^2/s}$. The figure shows that $D$ indeed depends on $k$, starting to decrease at a wavenumber around $k\approx \unit[10^7]{m^{-1}}$. This value is connected to a length scale $2\pi/k$ of roughly 600 nm, which is indeed a good estimate for a typical length of micellar particles \cite{buchanan2005high}.

As the particle is probing length scales comparable to the length scale of (nonlinear) interaction with the bath (recall that the length of worm-like micelles is between $\unit[100]{}$ and $\unit[1000]{nm}$ \cite{walker2001rheology}, and the typical mesh size is on the order of $\unit[30]{nm}$ \cite{buchanan2005high}), it apparently experiences a greater resistance which is reflected in a decrease of the diffusion coefficient. In the opposite regime of small wavenumbers, $k\to 0$, the continuous properties of the bulk system enter and the diffusion coefficient reaches its plateau value $D_0$. Note that resolving the long-time diffusion coefficient for high $k$ becomes more and more difficult due to a finite spatial accuracy of $\unit[4]{nm}$ in the experiment \cite{crocker1996methods}, and a sharper decay of the average in Eq.~\eqref{eq:diffcoeff_interacting} over time.

Having obtained a first indication of the pronounced nonlinear properties of the micellar bath without confinement, we next develop a theoretical understanding of nonlinear markers in the presence of a confining potential.
\section{Theory: How nonlinear interactions enter linear coefficients}\label{sec:theoretical_approach}
In this section we aim to analyze how nonlinear interactions can modify or enter linear coefficients. Using the well-known Zwanzig Mori projection operator technique, we obtain a linearized equation of motion for a single  (the colloidal) degree of freedom, coupled to bath particles making up the viscoelastic medium. This analysis will be performed in the two cases of Hamilton dynamics as well as overdamped dynamics.

\subsection{Hamilton dynamics}
As discovered by Mori in 1965 using projection operator techniques, the Liouville equation \cite{hansen2013theory}, describing the dynamics of a Hamiltonian system, can  be transformed into a linear equation for an observable  $A$ of interest (which can be vector in space of observables) \cite{mori1965transport,hynes1975nonequilibrium}
\begin{equation}\label{eq:Mori_GLE}
	\frac{\partial}{\partial t}A(t)=i\vct{\Omega}\cdot A(t)-\int_0^t \dd{s}{\bf M}(s)\cdot A(t-s)+F(t)\,,
\end{equation}
with the matrices (in space of observables) $\vct{\Omega}$ and ${\bf M}$ given by
\begin{align}
i\vct{\Omega}&=(\mathcal{L}A,A)\cdot(A,A)^{-1}\,,\\
{\bf M}(t)&=(F(t),F(0))\cdot(A,A)^{-1}\,.\label{eq:memkernel_Mori}
\end{align}
$\mathcal{L}$ is the Liouville operator and the parentheses $(\,\cdot\,,\,\cdot\,)$ specify an inner product weighted with the equilibrium distribution function $f_\mathrm{eq}$
\begin{equation}
(A,B)=\int \dd{X} f_\mathrm{eq}(X)A(X)B^*(X)=\langle A B^*\rangle_\mathrm{eq}\,.
\end{equation}
Note that the asterisk denotes the complex conjugate. The integral over $X\equiv(\vct{p},\vct{q})$ is meant to be over all phase space variables (here position and momentum degrees of freedom $\vct{q}$ and $\vct{p}$). The process of integrating out degrees of freedom to arrive at Eq.~\eqref{eq:Mori_GLE} renders it non-Markovian, and also gives rise to a random force  (noise) $F(t)$ . It is given by the projected dynamics
\begin{equation}\label{eq:noise_Mori}
F(t)=e^{t(1-\vct{P})\mathcal{L}}(\vct{1}-\vct{P})\mathcal{L}A\,,
\end{equation}
where the projector $\vct{P}$, projecting on $A$, has been introduced. Its action on a variable $B$ is

\begin{equation}\label{eq:proj_operator}
 \vct{P}B=(B,A)\cdot(A,A)^{-1}\cdot A\,.
\end{equation}
Notably, Eq.~\eqref{eq:memkernel_Mori} may be identified with the fluctuation-dissipation theorem, linking the memory matrix ${\bf M}(t)$ to the equilibrium noise correlator.

Applying this to the case of a colloidal particle in a complex bath, we start from the following microscopic Hamiltonian,
\begin{equation}
	H=\frac{p^2}{2m}+V_\mathrm{ext}(x)+\sum_{j=1}^N \frac{p_j^2}{2m_j}+V_\mathrm{int}(\{\xi_j\})\,.
	\label{eq:H}
\end{equation}
Here, $m$ is the mass of the colloid, and $V_\mathrm{ext}(x)$ is an external potential acting on it, e.g., imposed by optical forces as mentioned above. $m_j$ are the masses of the $N$-bath particles. $p$ and $p_j$ are corresponding momenta. The potential $V_\mathrm{int}(\{\xi_j\})$ is the interaction potential of the $N+1$ particles involved, which is not necessarily pairwise additive. Since $V_\mathrm{int}(\{\xi_j\})$ is invariant under displacing all particles by the same vector, it can be given in terms of $\{\xi_j\}$, where $\xi_j\equiv q_j-x$ is the distance between tracer and bath particle $j$. For simplicity, we consider a one-dimensional system, expecting the qualitative discussion to be equivalent in other dimensions.

Aiming to describe the dynamics of the colloid, we naturally choose $A=(x,p)^T$, i.e., the vector formed by its  position and momentum. 
Using $\vct{P}B=\langle B x\rangle \langle x^2\rangle^{-1}x+\langle B p\rangle\langle p^2\rangle^{-1} p$,
we obtain from Eq.~\eqref{eq:Mori_GLE} the explicit result \cite{zwanzig2001nonequilibrium}
\begin{align}
\dot{x}(t)&=p(t)/m\\
m\ddot{x}(t)&=- \kappa x(t)-\int_0^t \dd{s} \Gamma(s) \dot{x}(t-s)+F(t)\,.
\label{eq:Leq}
\end{align}
While the applied technique is well known, it is worth reminding that despite the fact that the Hamiltonian in Eq.~\eqref{eq:H} contains nonlinear interactions $V_\mathrm{ext}$ and $V_\mathrm{int}$, the resulting Eq.~\eqref{eq:Leq} is linear in $p$ and $x$. Indeed, the nonlinear character of $V_\mathrm{ext}$ and $V_\mathrm{int}$ finds its way into the linear coefficients appearing in Eq.~\eqref{eq:Leq}. First, an effective spring constant $\kappa$ appears \cite{zwanzig2001nonequilibrium}, 
\begin{align}\label{eq:kapef}
\kappa=\frac{k_B T}{\langle x^2\rangle^\mathrm{eq}}.
\end{align}
which depends on $V_\mathrm{ext}$ (via the equilibrium distribution $f_\mathrm{eq}$) and thermal energy $k_BT$. It is however independent of $V_\mathrm{int}$. The so-called {\it memory kernel} $\Gamma(t)$ reads
\begin{equation}
\Gamma(t)=m(e^{t(\vct{1}-\vct{P})\mathcal{L}}(\vct{1}-\vct{P})\mathcal{L}p,\mathcal{L}p)\,(p,p)^{-1}\,.
\label{eq:Gam}
\end{equation}
Compared to $\kappa$ in Eq.~\eqref{eq:kapef}, the form of $\Gamma$ in Eq.~\eqref{eq:Gam} is more involved, containing the projected dynamics, and no closed form for it is known. However, the series expansion in time $t$ can be given, in principle, to any order, yielding more insight. Writing this expansion
\begin{equation}\label{eq:mem_series_Mori}
	\Gamma(t)=\sum_{n=0}^\infty \frac{\Gamma^{(2n)}}{(2n)!}t^{2n}\,,
	\end{equation}
	we note that only even powers of $t$ contribute due to the intrinsic time reversal symmetry $\Gamma(t)=\Gamma(-t)$ seen in Eq.~\eqref{eq:memkernel_Mori}. By expanding the operator exponential in Eq.~\eqref{eq:Gam}, and using the anti-Hermitian property of the Liouville operator, $\mathcal{L}=-\mathcal{L}^\dagger$, the Taylor coefficients in Eq.~\eqref{eq:mem_series_Mori} are found to be given by the quadratic form 
\begin{equation}
\Gamma^{(2n)}=(-1)^n m\langle p^2\rangle_\mathrm{eq}^{-1}\langle[(\vct{1}-\vct{P})\mathcal{L}]^{n+1}p]^2\rangle_\mathrm{eq}\,.
\end{equation}
For simplicity, we shall in the following consider the case of $V_\mathrm{ext}(x)=\frac{1}{2} \kappa x^2$. 
We find for the first two coefficients (where $\beta=(k_B T)^{-1}$) 
\begin{align}
	\Gamma^{(0)}={}&\beta\langle F_\mathrm{int}^2\rangle_\mathrm{eq}\label{eq:1st}\\
\begin{split}
\Gamma^{(2)}={}&-\frac{1}{m}\sum_{j,k}\langle \partial_j F_\mathrm{int}\;\mathrm{;}\,\partial_k F_\mathrm{int}\rangle_\mathrm{eq}\\
&-\sum_j \frac{1}{m_j}\langle(\partial_j F_\mathrm{int})^2\rangle_\mathrm{eq}\,.\label{eq:2nd}
\end{split}
\end{align}
We introduced the covariance $\langle A\,\mathrm{;}\,B\rangle=\langle A B\rangle -\langle A\rangle\langle B\rangle$ in Eq.~\eqref{eq:2nd}, and denote $F_\mathrm{int}(\{\xi_j\})=\sum_j\partial_j V_\mathrm{int}(\{\xi_j\})$, the force acting on the tracer particle due to interactions with the bath particles. While the leading term for short times, Eq.~\eqref{eq:1st}, depends on this interaction potential in an expected manner, already the second term, Eq.~\eqref{eq:2nd}, is more interesting: The first term on the right-hand side of Eq.~\eqref{eq:2nd} depends on the mass $m$ of the colloidal particle. The presence of this term is worth noting, as it goes against a naive expectation that the friction kernel should only depend on properties of the bath, and be independent of tracer mass. Indeed, this term carries the covariance of force gradients as a prefactor, and is thus absent for harmonic couplings, as e.g.~employed in the models by Caldeira-Leggett \cite{caldeira1981influence}. Such dependence on the tracer mass is hence a signature of  nonlinear coupling to the bath.
Going one term further, we have 
\begin{align}\label{eq:Mori_t4}
\Gamma^{(4)}&=\frac{\kappa}{m^2}\sum_{j,k}\langle \partial_j F_\mathrm{int}\;\mathrm{;}\,\partial_k F_\mathrm{int}\rangle_\mathrm{eq} + \mathcal{O}(\kappa^0)\,.
\end{align}
This term shows the emergence of another interesting dependence: The friction kernel does not only depend on the properties of bath and tracer, but also on the stiffness  $\kappa$ of the surrounding potential. (Let us be reminded that this potential does not act on the bath particles in Eq.~\eqref{eq:H}, in contrast to the analysis provided in Ref.~\cite{lisy2019generalized}).  This dependence has indeed been observed in computer simulations of molecular solutes in water in Ref.~\cite{daldrop2017external}. Notably, that term does also depend on the mass of the tracer particle, and it vanishes for harmonic tracer-bath coupling. It is thus another marker for nonlinear interactions, being absent in linear processes. This can be seen also in the higher-order terms in $t$,  which take the form,  
\begin{align}
	\Gamma^{(2n)}&=(-1)^n\frac{\kappa^{n-1}}{m^{n}}\sum_{j,k}\langle \partial_j F_\mathrm{int}\;\mathrm{;}\,\partial_k F_\mathrm{int}\rangle_\mathrm{eq} + \mathcal{O}(\kappa^{n-2})\,.
\end{align}

\subsection{Overdamped dynamics}
A similar analysis as done in the previous subsection is feasible for the case of overdamped dynamics, where a simple model consists of two coupled Brownian particles, i.e.,
\begin{align}
\begin{split}\label{eq:EOM_sim}
	\gamma\dot{x}(t)&=-V_{\mathrm{int}}'(x-q)-\partial_{x} V_\mathrm{ext}(x(t))+F(t)\,,\\
\gamma_b\dot{q}(t)&=V_{\mathrm{int}}'(x-q)+F_b(t)\,.
\end{split}
\end{align}
Here the position of the tracer particle $x(t)$ (the colloid)  is confined by a potential, which, as above, we take harmonic,  $V_\mathrm{ext}=\frac{1}{2} \kappa x^2$, and it interacts via an arbitrary interaction potential $V_\mathrm{int}(x-q)$ with the second  Brownian particle (the bath particle).
The noise sources of tracer and bath particles are assumed to be white, Gaussian, and independent ($(i,j) \in \{F,F_b\}$)
\begin{equation}
\langle F_i(t)\rangle_\mathrm{eq}=0\,, \quad \langle F_i(t)F_j(t')\rangle_\mathrm{eq}=\delta_{ij} 2 k_B T\gamma_i\delta(t-t')\,.
\end{equation}
Similarly to the above, one may obtain a linearized equation of motion for the position of the tracer particle, which reads 
\begin{align}
0=- \kappa x(t)-\int_0^t \dd{s} \Gamma(s) \dot{x}(t-s)+F(t).
\label{eq:Seq}
\end{align}
In this case, the memory kernel takes on the following expansion
\begin{equation}\label{eq:mem_series_FP}
	\Gamma(t)=\gamma\delta(t)+\sum_{n=0}^\infty\frac{\Gamma^{(n)}}{n!}t^n,
\end{equation}
which differs from Eq.~\eqref{eq:mem_series_Mori}: In the overdamped case, $\Gamma(t)$ is a nonanalytic function of time, carrying an instantaneous response $\sim \delta(t)$, and, despite its time-symmetry, even and odd powers of $t$.
A straightforward calculation then yields the coefficients in Eq.~\eqref{eq:mem_series_FP} 
\begin{align}
\Gamma^{(0)}={}&\beta \langle F_{\mathrm{int}}^2\rangle_\mathrm{eq}\\
\begin{split}
\Gamma^{(1)}={}&-\frac{1}{\gamma}\langle F_\mathrm{int}^{(1)}\;\mathrm{;}\,F_\mathrm{int}^{(1)}\rangle_\mathrm{eq}\\
&+\frac{1}{\gamma_b}\left(\beta\langle F_\mathrm{int}^2F_\mathrm{int}^{(1)}\rangle_\mathrm{eq}+\langle F_\mathrm{int}F_\mathrm{int}^{(2)}\rangle_\mathrm{eq}\right)\label{eq:2ndo}
\end{split}\\
\Gamma^{(2)}={}&\frac{\kappa}{\gamma^2}\langle F_\mathrm{int}^{(1)}\;\mathrm{;}\,F_\mathrm{int}^{(1)}\rangle_\mathrm{eq}+\mathcal{O}(\kappa^0)\label{eq:kexto}\\
\vdots\notag\\
\Gamma^{(n)}={}&(-1)^n\frac{\kappa^{n-1}}{\gamma^n}\langle F_\mathrm{int}^{(1)}\;\mathrm{;}\,F_\mathrm{int}^{(1)}\rangle_\mathrm{eq}+\mathcal{O}(\kappa^{n-2}).
\end{align}
The noted dependence on colloidal mass $m$ in, e.g.,  Eq.~\eqref{eq:2nd} is here, in the overdamped case, mirrored by the dependence on $\gamma$ in  Eq.~\eqref{eq:2ndo}: It is worth noting that the memory kernel depends on the bare tracer friction $\gamma$ (and not only on bath properties). As is the case in Eq.~\eqref{eq:2nd}, the term involves the force variance, i.e., it vanishes for harmonic tracer-bath couplings. Also the appearance of $\kappa$ in Eq.~\eqref{eq:kexto} is similar as in Eq.~\eqref{eq:Mori_t4}, with the mass replaced by the bare friction coefficient. 
Again, in any higher order in $t$, $\kappa$ and $\gamma$ appear, in combination with the covariance of the derivative of the interaction force.   

\section{Exploring different tracer-bath couplings}\label{sec:exploring_couplings}
\subsection{Model}
\begin{figure}[t]
\includegraphics[width=.9\linewidth]{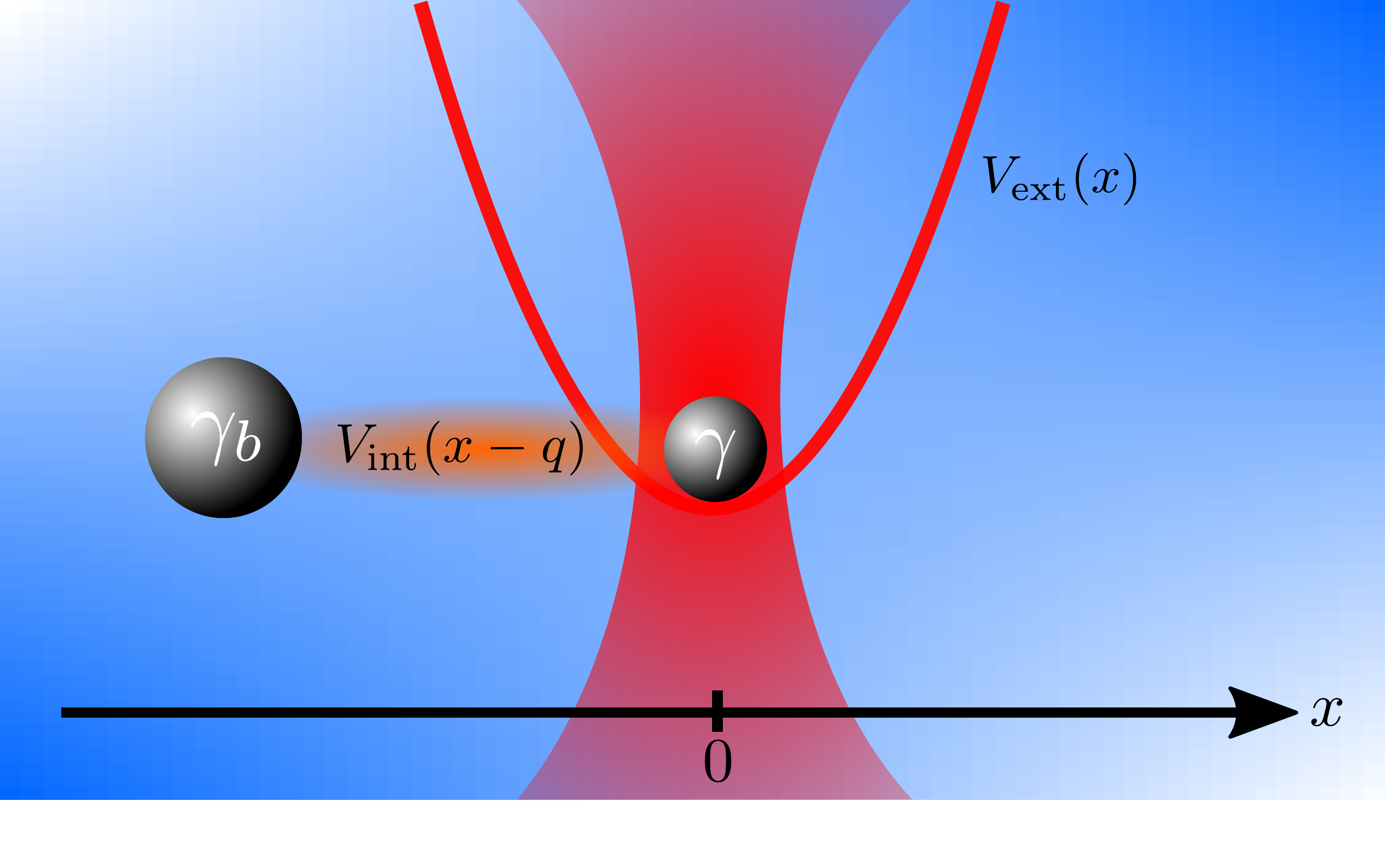}
\caption{\label{fig:model_sim}Model of tracer and bath. The tracer particle (coordinate $x$) is confined by a potential $V_\mathrm{ext}(x)=\frac{1}{2}\kappa x^2$, tracer and bath are coupled  via an interaction potential $V_\mathrm{int}(x-q)$, where $q$ is the coordinate of the bath particle.}
\end{figure}

Section \ref{sec:theoretical_approach} explicitly described the dependencies of the coefficients arising in a linearized equation; the friction memory of the bath may depend on the mass or the bare friction of the tracer, or the potential that the tracer is subjected to. Here we aim to study these using specific forms of $V_{\mathrm{int}}$ in Eq.~\eqref{eq:EOM_sim}, employing the model sketched in Fig.~\ref{fig:model_sim}: The (overdamped) colloidal particle is subject to a harmonic potential $V_\mathrm{ext}(x)=\frac{1}{2}\kappa x^2$. Additionally, the colloid is coupled to another overdamped particle, accounting for the bath. This model is thus given by Eq.~\eqref{eq:EOM_sim}, and is designed to mimic our experimental setup.

To compute the correlation function from simulated trajectories we deploy a stochastic Runge-Kutta method of weak convergence order three \cite{debrabant2010runge}. A high convergence order algorithm in combination with sufficient statistics is essential for our  analysis, as will be demonstrated in subsection \ref{sec:HC}.

While, from Eqs.~\eqref{eq:2ndo} and \eqref{eq:kexto}, interesting behavior of the friction kernel upon varying the colloidal bare friction $\gamma$ or $\kappa$ is expected, we will restrict ourselves to varying $\kappa$, as this is easily done in our experiments, and also allows comparison to results of Ref.~\cite{daldrop2017external}.

In the following, we set the friction coefficients of tracer and bath particle to be $\gamma=1$ and $\gamma_b=10$, respectively. Further, we consider inverse temperature $\beta=1$ and also all parameters appearing in the interaction potential $V_\mathrm{int}$ to unity. $\kappa$ can thus be thought of being given in units of $k_BT/d_0^2$ (see e.g. Eq.~\eqref{eq:Vint_PT_model} below for $d_0$), so that the length $\sqrt{\langle x^2\rangle_{\rm eq}}=\sqrt{k_BT/\kappa}$ is compared to the length scale $d_0$ appearing in the interaction potential.  The table of parameters used in this section are provided in Appendix \ref{app:par}. 

\subsection{Harmonic coupling}\label{sec:HC}
\begin{figure}[t]
\includegraphics[width=\linewidth]{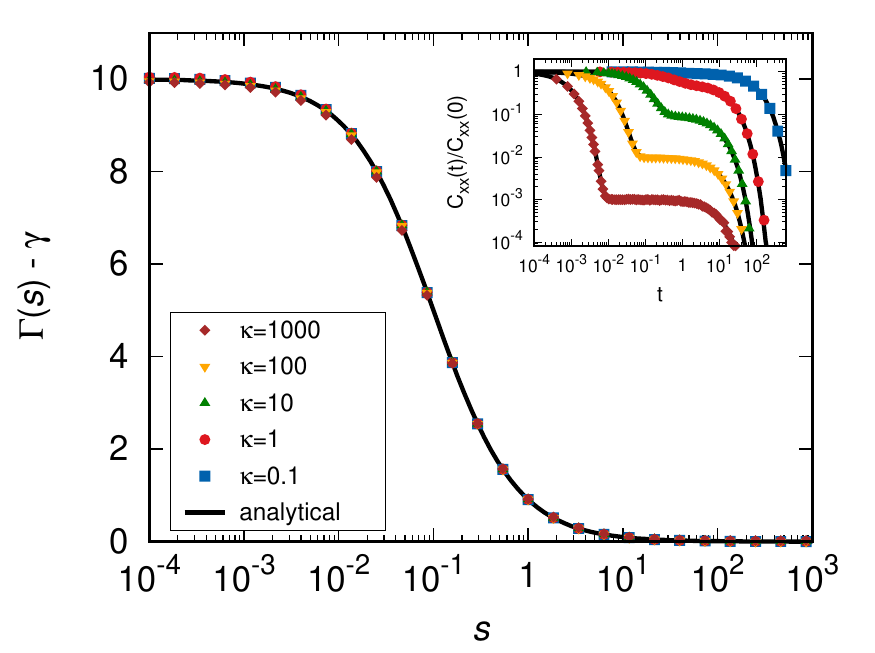}
\caption{\label{fig:mem_sim_kl}Main graph: Memory kernel $\hat{\Gamma}(s)-\gamma$ for harmonic coupling, in Laplace space, as obtained in numerical simulations, 
	for  different values of $\kappa$. The black line represents the analytical solution given in Eq.~\eqref{eq:memkernel_linear}. Inset: Simulated correlation function $\langle x(t)x(0)\rangle$ plotted against the analytical solution. The simulation parameters are provided in Table \ref{tab:mem_sim_kl}.}
\end{figure}
We start with the simplest case of a harmonic coupling, i.e., $V_\mathrm{int}(\xi)=\frac{1}{2}\kappa_l \xi^2$, which can be treated analytically \cite{zwanzig2001nonequilibrium,zwanzig1973nonlinear,caldeira1981influence}. One obtains the following linear equation for the colloidal particle 
\begin{equation}\label{eq:GLE_overdamped}
	\int_{-\infty}^t \dd{s} \Gamma(t-s)\dot{x}(s)=-\kappa x(t)+\tilde{F}(t)\,.
\end{equation}
The  kernel and noise are 
\begin{align}
\Gamma(t)&=2\gamma\delta(t)+\kappa_le^{-\frac{\kappa_l}{\gamma_b}t}\,,\label{eq:memkernel_linear}\\
\quad \tilde{F}(t)&=F(t)+\frac{\kappa_l}{\gamma_b}\int_{-\infty}^t \dd{s}e^{-\frac{\kappa_l}{\gamma_b}(t-s)}F_b(s)\,.
\end{align}
As expected from section \ref{sec:theoretical_approach} and literature \cite{zwanzig2001nonequilibrium,zwanzig1973nonlinear,caldeira1981influence}, the kernel $\Gamma(t)$ in Eq.~\eqref{eq:memkernel_linear}, apart from the trivial term $2\gamma\delta(t)$ depends only on bath properties, i.e., $\gamma_b$ and the interaction strength $\kappa_l$.

The harmonic coupling allows to test the quality of numerical simulations, applied to Eqs.~\eqref{eq:EOM_sim}. 
These are used to create particle trajectories, from which the correlation function $C_{xx}(t)=\langle x(t) x(0)\rangle_\mathrm{eq}$ is computed.  Aiming to extract $\Gamma(t)$, we turn to Laplace space, where Eq.~\eqref{eq:GLE_overdamped} reads 
\begin{equation}\label{eq:memkernel_LT}
\hat{\Gamma}(s)=\frac{\kappa \hat{C}_{xx}(s)}{\frac{k_BT}{\kappa}-\hat{C}_{xx}(s) s}
\end{equation}
with Laplace transforms $\hat{h}(s)=\int_0^\infty \dd t e^{-st}h(t)$. Note that we used in this derivation the equal time correlation function to be given by $C_{xx}(0)=\langle x^2\rangle_\mathrm{eq} =\frac{k_B T}{\kappa}$, i.e., the bath is prepared in thermal equilibrium at $t=0$. 

Figure \ref{fig:mem_sim_kl} shows the results for different values of $\kappa$. The data points for $\Gamma(s)$ (main graph)  and $C_{xx}(t)$ (inset) follow well the analytical forms (solid lines). For values of $\kappa$ spanning five orders of magnitude,  $\Gamma(s)$, as found from simulations, takes identical forms. This requires high numerical accuracy, as may be illustrated by regarding the case of $\kappa=10^3$: Here, the correlator quickly decays to a the very small value of $\sim 10^{-3}$, reaching a plateau value there. This plateau, which is easily overlooked, is however essential to obtain the correct value of $C_{xx}(s)$ as $s\to0$.  In the limit of large $\kappa$, the relaxation time scales set by the trap and the bath are well separated and the latter is decisive for the Laplace transform of the correlation function. Note that in the limit of large $s$ (small times) the particle diffuses freely, i.e.\ we have $\lim_{s\to \infty}\hat{\Gamma}(s)=\gamma$, while for small $s$ (large times) $\hat{\Gamma}(s=0)=\gamma+\gamma_b$ due to the bounded potential.

\subsection{Double-well interaction potential}\label{sec:DW}

\begin{figure}[t]
\includegraphics[width=\linewidth]{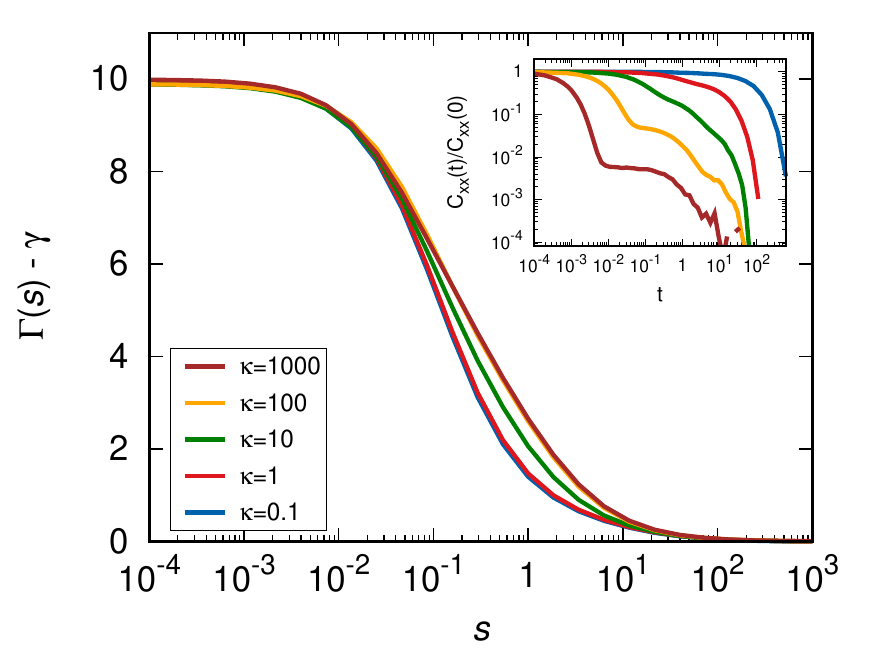}
\caption{\label{fig:mem_sim_dw}Main graph: Memory kernel $\hat{\Gamma}(s)-\gamma$ for a double well coupling, in Laplace space, as obtained in numerical simulations, for various values of  $\kappa$. Inset: Simulated correlation function $\langle x(t)x(0)\rangle$. The simulation parameters are provided in Table \ref{tab:mem_sim_dw}.}
\end{figure}

Going one step beyond subsection \ref{sec:HC}, we consider a nonlinear interaction potential. A useful choice for such nonlinear potential is a symmetric double well 
\begin{equation}
V_\mathrm{int}(\xi)=\frac{V_0}{d_0^4}(\xi-d_0)^2(\xi+d_0)^2\,,
\end{equation}
where  $V_0$ is the height of the potential barrier between the wells, and $d_0$ is half the distance between their minima. Note that for a single particle moving in such a potential interesting barrier-crossing kinetics were found for a Langevin equation with bi-exponential memory only recently \cite{kappler2019non}.

The results for the Laplace transformed memory kernel $\hat{\Gamma}(s)$ are illustrated in Fig.~\eqref{fig:mem_sim_dw}. In contrast to the case of harmonic coupling, the memory kernel shows indeed, as expected from Eq.~\eqref{eq:kexto}, a dependence on the external trap stiffness $\kappa$ for finite values of $s$. However, for small and large values of $s$, $\hat{\Gamma}(s)$ takes the same limiting values as for the harmonic coupling, shown in Fig.~\ref{fig:mem_sim_kl}. For large $s$, the colloid, as before, does not notice the presence of the bath. For small $s$, the two particles behave as a composite particle, with the sum of bare friction coefficients. This is because the two particles are coupled by a bound potential, so that on large time scales, they move together.

\subsection{Stochastic Prandtl-Tomlinson model}
\begin{figure}[t]
\includegraphics[width=\linewidth]{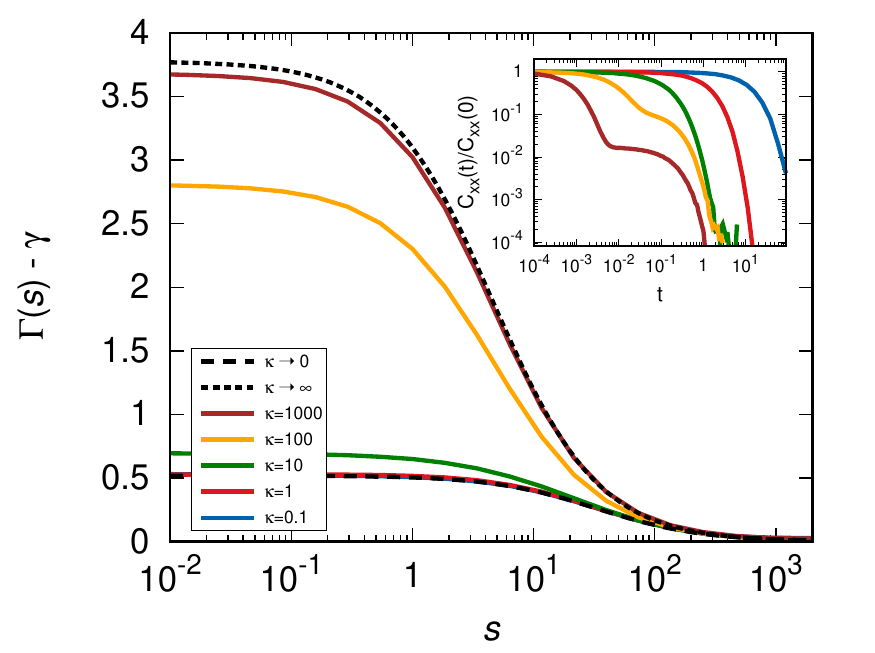}
\caption{\label{fig:mem_sim_osc}Main graph: Memory kernel $\hat{\Gamma}(s)-\gamma$ for the stochastic Prandtl-Tomlinson model, in Laplace space, as obtained in numerical simulations, for various values of  $\kappa$. Inset: Simulated correlation function $\langle x(t)x(0)\rangle$. The limiting curves for $\kappa\to0$ and $\kappa\to\infty $ shown in the main graph will be discussed in Sec.~\ref{sec:limiting_cases}. The simulation parameters are provided in Table \ref{tab:mem_sim_pt}}
\end{figure}

Given the observation of subsection \ref{sec:DW}, we now look for a model where colloid and bath particle are not bound, which brings us to the so called Prandtl-Tomlinson (PT) model. This model is popular  in the field of frictional processes on the atomic scale, and was introduced by  Prandtl to describe plastic deformations in crystals as well as dry friction \cite{prandtl1928gedankenmodell}. The model consists of a particle in a periodic potential that is damped by a frictional force and harmonically coupled to a host solid. The theoretical framework derived by Prandtl has been extensively deployed and modified to be applicable to a wide range of physical applications \cite{popov2012prandtl}. Most prominently it well describes the damped motion of a nanotip of an atomic force microscope driven over a (corrugated) surface \cite{gnecco2000velocity,jansen2010temperature,mueser2011velocity}.

Using \begin{equation}\label{eq:Vint_PT_model}
V_\mathrm{int}(\xi)=V_0\cos(\frac{2\pi}{d_0}\xi)\,,
\end{equation}
 with a wavelength $d_0$ and amplitude $V_0$, with Eq.~\eqref{eq:EOM_sim} yields an extension of the PT model: It is a PT model where the sinusoidal potential is not fixed in space, but is the interaction potential with a bath particle, which by itself is  stochastic  with finite friction and diffusion coefficients (given through $\gamma_b$).  The physical intuition  is that the micellar bath of our experiments is indeed a non-static background with a finite relaxation time.

In contrast to subsection \ref{sec:DW}, $V_\mathrm{int}$ in Eq.~\eqref{eq:Vint_PT_model}  is  unbounded. 
Figure \ref{fig:mem_sim_osc} shows the results obtained from simulations of the stochastic PT model for different values of $\kappa$. As expected,  $\hat{\Gamma}(s)$ does strongly depend on $\kappa$, and, as the potential is unbound, also in the limit of $s\to0$.  The value of $\hat{\Gamma}(s=0)$ differs by more than a factor of three between very large and very small values of $\kappa$ (a ratio which can of course be tuned by varying the model parameters).

\begin{figure}[t]
\includegraphics[width=\linewidth]{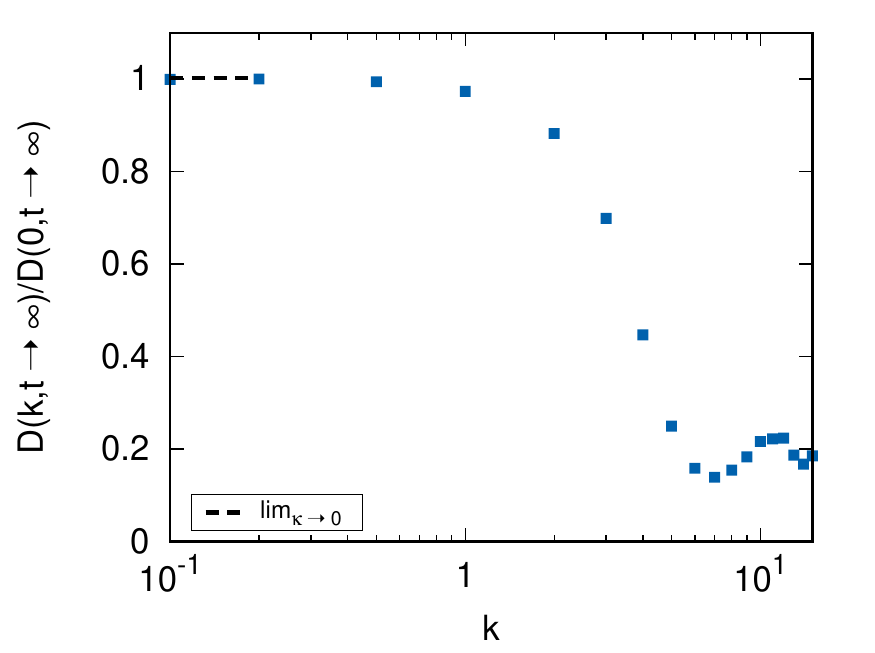}
\caption{\label{fig:long_time_diffusion_sim}Normalized wavenumber-dependent long-time diffusion coefficient $D(k,t\to\infty)$ extracted from a free particle simulation of the stochastic PT model.}
\end{figure}

Can the stochastic PT model describe our experiments? Before addressing this question quantitatively in section \ref{sec:periodic_coupling_model} below, we first study the diffusion coefficient $D(k,t)$, obtained as in Fig.~\ref{fig:diff_coeff_long_time_exp}, but here from simulation trajectories performed at $\kappa=0$. Figure \ref{fig:long_time_diffusion_sim} shows the resulting values as a function of wavevector $k$ (see Eq.~\eqref{eq:diffcoeff_interacting}). As was observed in Fig.~\ref{fig:diff_coeff_long_time_exp}, the data points Fig.~\ref{fig:long_time_diffusion_sim} decrease with increasing $k$, while they approach limiting values for both large and small $k$. While in Fig.~\ref{fig:diff_coeff_long_time_exp} the characteristic wavevector was identified with the size of micellar particles, it is here related to the chosen value of $d_0$ (which is unity). 

Encouraged by the qualitative agreement between our experiments and the stochastic PR model, we continue with a detailed and  quantitative comparison in the next section.

\section{\label{sec:periodic_coupling_model}Stochastic Prandtl-Tomlinson model and experiments}
Having discussed several signatures of bath-nonlinearity in equilibrium systems, we now aim to discuss these in quantitative detail for our experiments, in comparison with the stochastic PT model.

\begin{figure}[t]
\includegraphics[width=\linewidth]{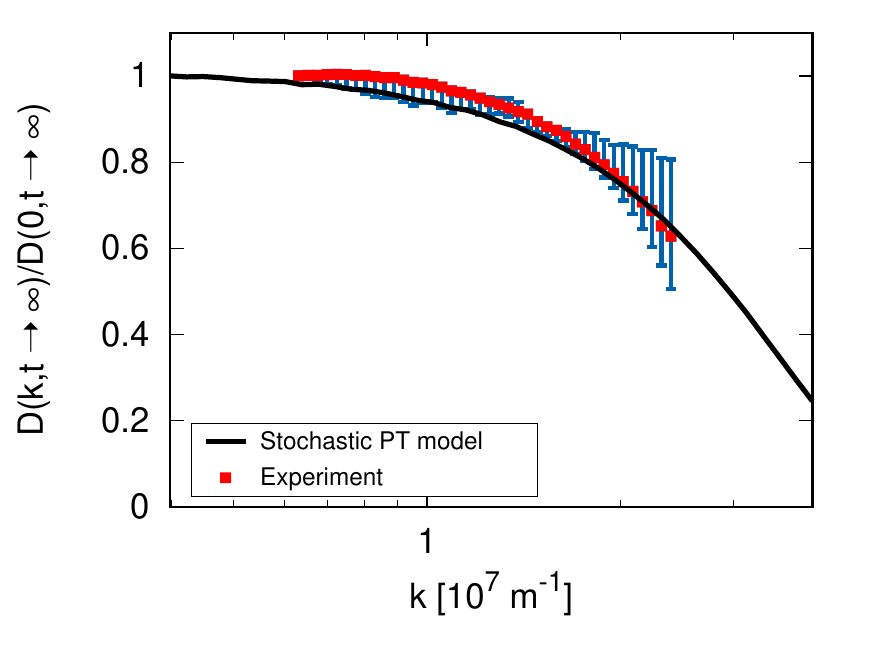}
\caption{\label{fig:free_diff_fit}Normalized wavenumber-dependent long-time diffusion coefficient $D(k,t\to\infty)$ for a micellar bath, obtained from experiments as well as from the stochastic PT model. As in Fig.~\ref{fig:diff_coeff_long_time_exp} above, bars show the statistical error estimated by partitioning the measured trajectory into two pieces.}
\end{figure}

Starting with the diffusion coefficients, Fig.~\ref{fig:free_diff_fit} shows the experimental data of Fig.~\ref{fig:diff_coeff_long_time_exp} together with those obtained from the SPT model, showing good agreement. The fit parameters of the model are provided in Table \ref{tab:mem_exp}.

\begin{figure}[b]
\includegraphics[width=\linewidth]{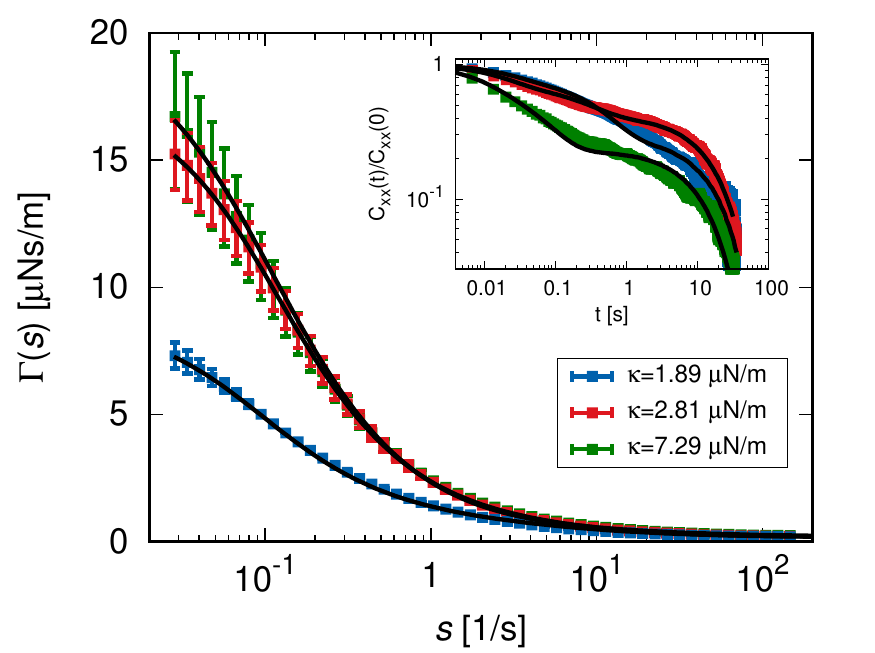}
\caption{\label{fig:mem_exp}Main graph: Memory kernel $\hat{\Gamma}(s)$ of a micellar system, in Laplace space, as obtained from experimental data, for three values of $\kappa$. Error bars show the statistical error obtained from partitioning the measured trajectories into two pieces.  Inset: Correlation function $\langle x(t)x(0)\rangle_\mathrm{eq}$ from experimental data. In both main graph and inset, the solid black lines are obtained from the stochastic PT model, with parameters given in table \ref{tab:mem_exp}.}
\end{figure}

Notably, the SPT model can be used to link our experimental data of free diffusion to the cases with optical trap present;
Fig.~\ref{fig:mem_exp} finally shows the experimental correlation functions $\langle x(t) x(0) \rangle_\mathrm{eq}$ for the trapped particle, at three different values of trapping strength $\kappa$ (inset). The main graph gives, in the same manner as Figs.~\ref{fig:mem_sim_kl}-\ref{fig:mem_sim_osc}, the extracted form of the friction memory $\Gamma(s)$. We  restrict to the experimentally accessible range of $s$. Indeed, our experiments also strongly show the marker of nonlinearity exhibited in $\Gamma$, as this function displays a dependence on the given $\kappa$.

The solid lines shown in Fig.~\ref{fig:mem_exp} have been obtained from the SPT model, with parameters given in Table \ref{tab:mem_exp}. The agreement with the data is very good, underpinning our conclusion that the effects seen in Figs.~\ref{fig:free_diff_fit} and \ref{fig:mem_exp} have the same physical origin: Nonlinear interactions on the micellar length scale.

Ideally, one set of parameters of the SPT model should suffice to describe all experimental data shown in Figs.~\ref{fig:free_diff_fit} and \ref{fig:mem_exp}. We have, however, slightly adjusted the parameters to obtain optimal agreement. Additionally to the circumstance that the stochastic PT model is a rather coarse representation of the micellar bath, there are also possibilities for systematic errors causing these parameter variations: Optical traps as used in our experiments are usually optimized for trap stiffnesses that are smaller than the largest ones used here, so that studying the mentioned effects requires going beyond the typical regime of stiffnesses. This may introduce local heating, promotion of ageing effects, or a slight anharmonicity of trap shape; Indeed, the relative standard error for $\kappa$ is roughly $2\%$ for its smallest value and $5\%$ for the largest. With these comments in mind, the quantitative agreement between experimental data and the stochastic PT model is satisfactory and convincing.

We also point out that the used version of SPT model cannot correctly account for the prefactor $D(0,t\to\infty)$ used to normalize the data  in Fig.~\ref{fig:free_diff_fit}. Indeed, Fig.~\ref{fig:free_diff_fit} shows only a small range of $k$, and processes on other length scales may influence $D(0,t\to\infty)$. The observation of the missmatch of $D(0,t\to\infty)$ suggests the presence of other important length scales, which theoretically could be accounted for by adding more bath particles with different values of $d_0$. In the absence of more experimental evidence at present, we leave this discussion for future work. We note that adding another bath particle with a distinct  length scale would not change the curves shown in Fig.~\ref{fig:mem_exp}.

The simulation parameters provided in Table \ref{tab:mem_exp} may now be interpreted in terms of experimental scales. The amplitude $V_0$ is a typically potential barrier formed by micelles surrounding the tracer particle, and its value being of order of $k_BT$ is thus reasonable.  The length scale $d_0$ sets the dominant length scale of (nonlinear) interactions between tracer particle and bath. It is here of the order of a few hundred nanometers, which is in great agreement with sizes of micellar particles \cite{buchanan2005high}. Finally, the relaxation time of the bath for a fixed tracer position can be estimated from $\gamma_b$ and the curvature of $V_{\rm int}$ at its minimum. It is of the order of a few seconds for the parameters of Table \ref{tab:mem_exp},  matching the order of magnitude of the measured structural relaxation time by a recoil experiment \cite{gomez2015transient}. The STP model thus well describes our experiments, with physically plausible parameters.

\begin{table}[]
\begin{ruledtabular}
\begin{tabular}{|l|l|l|l|l|}\hline
$\kappa$ [$\mu$N/m] & $V_0$ [$k_B T$] & $d_0$ [nm] & $\gamma$ [$\mu$Ns/m] & $\gamma_b$ [$\mu$Ns/m] \\\hline
0 & 2.1 & 98 & 0.16 & 148\\\hline
1.89 & 1.9 & 210 &  0.18 & 66.7 \\\hline
2.81 & 2.11 & 210 & 0.168 & 68.2 \\\hline
7.29 & 1.4 & 120 &   0.189 & 148.3\\\hline
\end{tabular}
\caption{\label{tab:mem_exp}Parameters of the stochastic PT model used for the curves shown in Figs.~\ref{fig:free_diff_fit} and \ref{fig:mem_exp}.}
\end{ruledtabular}
\end{table}

\section{Connection to micro-rheology and limiting cases}\label{sec:limiting_cases}
In this section, we establish ties between the results from previous sections and the viscosity or friction coefficient obtained from microrheology \cite{cicuta2007microrheology,gazuz2009active,squires2010fluid,wilson2011microrheology,harrer2012force,puertas2014microrheology,berret2016local}. This will also allow to better understand the limiting curves of $\Gamma(s)$ for very large and very small values of $\kappa$, as e.g. shown in Fig.~\ref{fig:mem_sim_osc}. 

In a typical setting of (active) microrheology, the potential trap moves at a constant velocity $v_0$, switched on at time $t=0$. The external potential is thus given by
\begin{equation}
V_\mathrm{ext}=\frac{1}{2}\kappa(x-v_0 t)^2\,.
\end{equation} 
Due to the motion of the trap, the colloidal particle is dragged through the bath, thereby giving rise to a friction force. The mean force exerted by the particle on the external potential (or vice versa), $\kappa|\langle x\rangle (t)-v_0 t|$, is thus the observable of interest. Division by $v_0$ yields the time dependent friction coefficient $\gamma(t)$. Using linear response theory (see Appendix \ref{app:lin_response_v}) $\gamma(t)$ may be connected to fluctuations of the colloid measured at rest,
\begin{equation}\label{eq:lin_response_v}
\gamma(t)\equiv \frac{\kappa|\langle x\rangle (t)-v_0 t|}{v_0}=\beta\kappa^2\int_0^t\dd{t'}\langle x(t')x(0)\rangle_\mathrm{eq}\,.
\end{equation}
Performing a  small $s$ expansion of $\hat{C}_{xx}(s)$ in Eq.~\eqref{eq:memkernel_LT}, we find a connection between the long-time friction coefficient of microrheology  and the memory kernel defined in the generalized Langevin equation, given by
\begin{equation}\label{eq:friction_coeff_long_time}
\gamma\equiv\lim_{t\to \infty}\gamma(t)=\hat{\Gamma}(0)=\int_0^\infty\dd{t'} \Gamma(t')\,.
\end{equation}
This relation provides an insightful connection: The previously discussed and analyzed form of $\Gamma(s)$, i.e., its dependence on parameters such as the colloidal mass or the trapping potential coefficient $\kappa$, thus translate to microrheological observations. Indeed, it has been noticed before, that, e.g., the value of the trapping potential coefficient $\kappa$ can influence the measured microrheological viscosity \cite{bishop2004optical,squires2005simple,brau2007passive,yao2009microrheology}.  

This insight becomes even stronger when discussing the limit of large and small $\kappa$. Taking the limit of ${\kappa\to\infty}$ allows neglecting the term on the left hand side of Eq.~\eqref{eq:GLE_overdamped}. Using this, as well as $\langle f(t)f(0)\rangle_\mathrm{eq}=\beta^{-1}\Gamma(|t|)$ and Eq.~\eqref{eq:lin_response_v}, we obtain, in the limit $\kappa\to\infty$, 
\begin{equation}\label{eq:mem_kext_inf}
\lim_{\kappa\to\infty}\hat{\Gamma}(s)=\lim_{\kappa\to\infty}s \hat{\gamma}(s)\,.
\end{equation}
Eq.~\eqref{eq:mem_kext_inf} is the extension of Eq.~\eqref{eq:friction_coeff_long_time} for any $s$, valid for large $\kappa$. The  microrheological setup corresponding to the limit $\kappa\to\infty$ is the case where the colloidal particle is moving at constant velocity $v_0$.  The dashed line in Fig.~\ref{fig:mem_sim_osc}, giving the limit for large $\kappa$, thus corresponds to the microrheological case of driving at constant velocity. In our simulations, we obtained it by measuring the correlator of forces acting on the tracer held at fixed position.

In order to address the reverse limit of vanishing trap stiffness, i.e., ${\kappa\to 0}$, we introduce the particle mobility $\mu(t)$ via the Einstein relation, 
\begin{equation}\label{eq:Einstein_relation}
\mu(t)\equiv\frac{\beta}{2}\dv{t}\langle (x(t)-x(0))^2\rangle_\mathrm{eq}\,.
\end{equation}
Expressing the right hand side in terms of the time derivative of $C_{xx}(t)$, we obtain from Eq.~\eqref{eq:memkernel_LT} an equation for $\mu(s)$
\begin{equation}
\hat{\mu}(s)=\frac{1}{s\hat{\Gamma}(s)+\kappa}=\frac{1}{s\hat{\Gamma}(s)}+\mathcal{O}(\kappa)\,.
\end{equation}
In the second equality we expanded for small $\kappa$. (Note that the limits $s\to 0$ and  $\kappa\to 0$ do not commute). 
We thus find
\begin{equation}\label{eq:mem_kext_0}
\lim_{\kappa\to 0}\hat{\Gamma}(s)=\lim_{\kappa\to 0}\frac{1}{s\hat{\mu}(s)}\,.
\end{equation}
We thus connected the limit of vanishing trap stiffness $\kappa$ to the mobility $\mu$ of the particle in absence of the trap. In the language of microrheology, this mobility is found by applying a constant force to the tracer. The second limiting curve in Fig.~\ref{fig:mem_sim_osc} corresponds thus to the microrheological mobility of constant applied force. The curve in the figure was obtained by measuring the diffusion process of of the tracer in absence of external potential. 

\begin{figure}[t]
 \includegraphics[width=\linewidth]{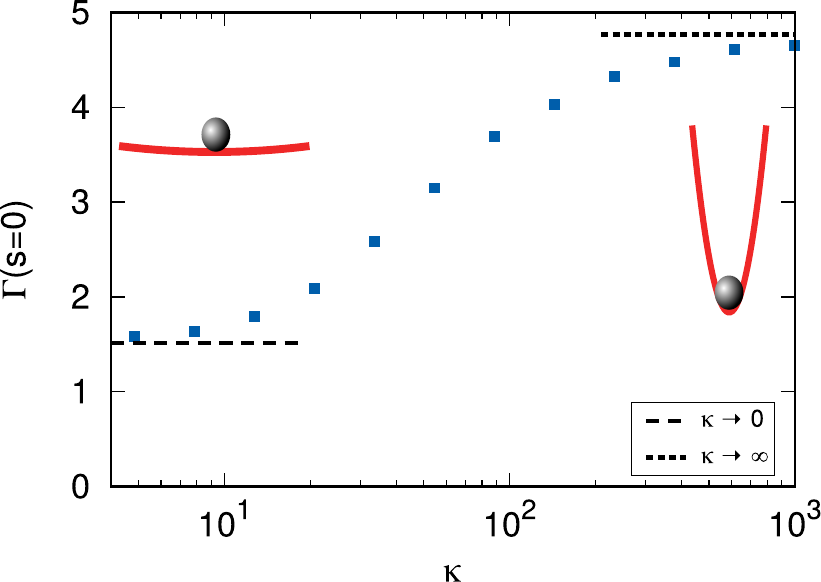}
 \caption{\label{fig:gm0_PT}Long-time friction coefficient $\gamma\equiv \lim_{t\to\infty}\gamma(t)=\hat{\Gamma}(0)=\int_0^\infty \dd{t'}\Gamma(t')$ of the particle in a slowly moving trap from the stochastic PT model. Dashed lines give the limiting cases $\kappa\to 0$ and $\kappa\to\infty$ as discussed in the main text. The simulation parameters are provided in Table \ref{tab:mem_sim_pt}.}
\end{figure}

Expanding Eq.~\eqref{eq:diffcoeff_interacting} for small $k$ and comparing to Eq.~\eqref{eq:Einstein_relation} yields the well-known relation
\begin{equation}
\lim_{\kappa\to 0} \hat{\Gamma}(0)=\frac{1}{\beta D(0,t\to\infty)}.
\end{equation}
Fig.~\ref{fig:long_time_diffusion_sim}, where we included the corresponding value taken from the limiting curve in Fig.~\ref{fig:mem_sim_osc} as a dashed line, displays this relation. No such relation has been obtained in the opposite limit for $k\to\infty$. 

We thus conclude that the measurement of linear response coefficients $\lim_{\kappa\to\infty}\gamma(t)$ and $\lim_{\kappa\to 0}\mu(t)$ gives  information on the $\kappa$-dependence of the memory kernel $\Gamma(t)$ if the effective dynamics of the system is modeled by a linear generalized Langevin equation. 

Note that a related discussion of linear response coefficients was presented in Ref.~\cite{squires2005simple} for a system with hard-sphere interactions. The authors analyze the linear response coefficients of the two extreme modes (constant force and constant velocity) and demonstrate their difference. Here, however, we aimed at a connection between the linear response coefficients and the friction memory kernel $\Gamma$ as frequently used in the description of effective Brownian dynamics. 

Figure \ref{fig:gm0_PT} shows $\Gamma(s=0)$ as a function of $\kappa$, obtained in the stochastic Prandtl Tomlinson model, including the two limiting cases for large and small $\kappa$.

\section{Summary}
Combining experimental measurements, analytical computations, and simulations of a stochastic Prandtl-Tomlinson model, we investigated several equilibrium-properties of a colloidal particle suspended in a nonlinear bath. Additionally to displacements being non-Gaussian, a nonlinear bath shows up by unexpected properties of the coefficients of a linearized equation; For example, in a nonlinear bath, the effective friction memory of the bath can depend on the stiffness of a potential trapping the particle, as has been observed in molecular simulations \cite{daldrop2017external}, or on the mass or bare friction of the colloidal particle. These dependencies are observed in our experiments, so that the friction memory of the trapped particle varies by more than a factor of two for trap stiffnesses $\kappa$ ranging from $\unit[1.89]{\mu N/m}\dots\unit[7.29]{\mu N/m}$. The mentioned dependencies are also demonstrated in an exact analytic expansion of the memory kernel for small times, for the case of Hamilton and Brownian dynamics.

Linear response theory provides the link between the measurements of the particle in the trap at rest and microrheological quantities. This allows to determine the limiting forms for the friction kernel for small and large trapping stiffness, and also connects this discussion to microrheological cases of ``constant force'' or ``constant velocity''. 

Analyzing several models, we develop a stochastic Prandtl-Tomlinson, which is easy to be evaluated numerically and which well describes all aspects of our experimental data. The resulting parameters of the model are physically plausible, so that this model promises to be useful in analysis of complex tracer-baths systems.

The observed ``fluctuation renormalisation'' of linear coefficients, already pointed out by Zwanzig, must be taken into account when describing nonlinear stochastic systems, and are pronounced for colloidal motion in viscoelastic solvents. External forces, inertial forces, and forces from a bath, which are strictly independent on a microscopic level, become dependent on each other in a linearized description.

Colloidal motion in viscoelastic baths, here exemplified by a micellar suspension, provide a new paradigmatic case of stochastic motion, with various phenomena that go far beyond the well-studied cases of colloids in pure solvents. It is thus important to develop basic understanding of such systems, and the studies performed here provide a first step towards systematic investigation and modeling. With the given findings at hand, future work will address non-equilibrium cases, and investigate how the equilibrium observations and modeling will determine nonlinear responses and far from equilibrium properties, such as those found in Refs.~\cite{gomez2015transient, berner2018oscillating}.

\begin{acknowledgments}
M.~K.~and B.~M.~were supported by DFG Grant No.~KR 3844/3-1. M.~K. and B.~M. also acknowledge support by the G\"ottingen Campus QPlus program. C.B. acknowledges financial support by the ERC Advanced Grant ASCIR (Grant No.693683) and from the German Research Foundation (DFG) through the priority program SPP 1726.
\end{acknowledgments}

\begin{appendix}
\section{\label{app:LT}Extraction of Laplace-transformed memory kernel and numerical error}
To extract the Laplace-transformed memory kernel $\hat{\Gamma}(s)$ from numerical (experimental or simulated) data, we use its unique relation to the position autocorrelation function $\hat{C}_{xx}(s)$ in Laplace domain given by Eq.~\eqref{eq:memkernel_LT} in the main text. We restrict our analysis of Laplace transforms to the real axis in Laplace domain, i.e.~we set $\Im[s]=0$.
\begin{figure}
\includegraphics[width=\linewidth]{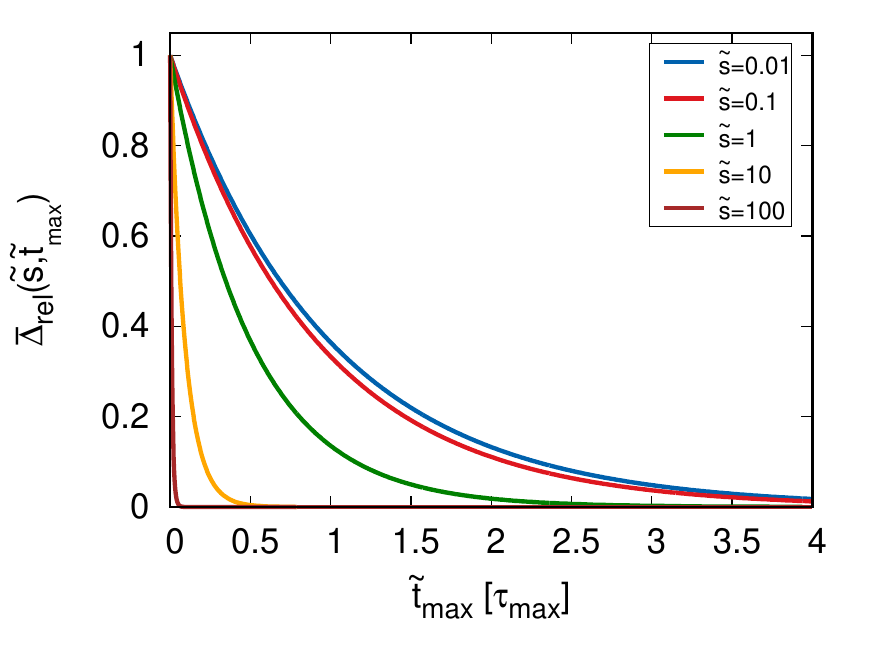}
\caption{\label{fig:num_approx}Relative numerical truncation error $\bar{\Delta}_\mathrm{rel}(\tilde{s},\tilde{t}_\mathrm{max})$ in rescaled units for a finite upper integration limit $\tilde{t}_\mathrm{max}$ in the Laplace transform of an exponentially decaying function $f(\tilde{t})=e^{-\tilde{t}}$ over a large range of $\tilde{s}$ values.}
\end{figure}
The Laplace transform of the equilibrium correlation function $\hat{C}_{xx}(s)=\int_0^\infty \dd{t}e^{-st}C_{xx}(t)$ is computed numerically up to a truncation time $t_\mathrm{max}$.
The equilibrium correlation function can be typically approximated by a sum of exponentially decaying function and thus we can approximate the relative numerical truncation error by
\begin{equation}\label{eq:num_truncation_err}
\Delta_\mathrm{rel}(s,t_\mathrm{max})=1-\frac{\int_0^{t_\mathrm{max}}e^{-st}f(t)\dd{t}}{\int_0^\infty e^{-st}f(t)\dd{t}}\,,
\end{equation}
where $f(t)=e^{-t/\tau_\mathrm{max}}$ is now the exponential decay with the longest relaxation time $\tau_\mathrm{max}$. By introducing rescaled variables $\tilde{s}=s\tau_\mathrm{max}$ and $\tilde{t}_\mathrm{max}=t_\mathrm{max}/\tau$ and computing the integrals in Eq.~\eqref{eq:num_truncation_err} analytically, the relative numerical truncation error takes on a form that does not explicitly depend on $\tau_\mathrm{max}$
\begin{equation}
\bar{\Delta}_\mathrm{rel}(\tilde{s},\tilde{t}_\mathrm{max})=\exp(-\frac{\tilde{t}_\mathrm{max}}{(1+\tilde{s})^{-1}})\,.
\end{equation}
The relaxation time of the numerical truncation error in rescaled units is bounded in the interval $[0,1]$ for any positive real value of $\tilde{s}$. The important time scales in the studied system under overdamped conditions are set by the relaxation time in the trap $\tau_\kappa=\gamma/\kappa$ and the relaxation time(s) of the memory kernel, which in the experiment is typically given by the structural relaxation time $\tau_s$. The relative numerical truncation error converges exponentially to zero as shown in Fig.~\ref{fig:num_approx} and is negligibly small even for small values of $\tilde{s}$ provided that $\tilde{t}_\mathrm{max}$ is a multiple of the longest relaxation time of the system. While the numerical truncation error is of negligible significance for simulated data (very long trajectories can easily be created), it might play a role for experimental data as the measurement time is limited e.g.~due to aging effects of the solution. 

\section{\label{app:par} Simulation parameters}
In this appendix, we provide the simulation parameters used for creation of the figures in section \ref{sec:exploring_couplings} and \ref{sec:limiting_cases} not given in the main text.

\subsection{Harmonic coupling}

\begin{table}[H]
\begin{ruledtabular}
\begin{tabular}{|l|l|l|l|l|l|l|l|l|}\hline
$\kappa$ & $n_\mathrm{traj}$ & $n_\mathrm{step}$ & $dt$ & $\beta$ & $\gamma$ & $\gamma_b$ & $\kappa_l$  & $t_\mathrm{eq}$\\\hline
$0.1$ & $1.8 \cdot 10^6$ & $10^6$ & $10^{-2.5}$ & $1$ & $1$ & $10$ & $1$ & $300$\\\hline
$1$ & $1.8 \cdot 10^6$ & $10^6$ & $10^{-2.7}$ & $1$ & $1$ & $10$ & $1$ & $100$\\\hline
$10$ & $1.8 \cdot 10^6$ & $10^6$ & $10^{-2.9}$ & $1$ & $1$ & $10$ & $1$ & $60$\\\hline
$100$ & $1.8 \cdot 10^6$ & $10^6$ & $10^{-3.4}$ & $1$ & $1$ & $10$ & $1$ & $40$\\\hline
$1000$ & $1.8 \cdot 10^6$ & $10^6$ & $10^{-3.4}$ & $1$ & $1$ & $10$ & $1$ & $20$\\\hline
\end{tabular}
\caption{\label{tab:mem_sim_kl}Simulation parameters for the case of harmonic coupling. $n_\mathrm{traj}$ denotes the number of trajectories, $n_\mathrm{step}$ is the number of time steps $dt$ of a single trajectory, and $t_\mathrm{eq}$ is the equilibration time used in the simulation.}
\end{ruledtabular}
\end{table}

\subsection{Double-well interaction potential}

\begin{table}[H]
\begin{ruledtabular}
\begin{tabular}{|l|l|l|l|l|l|l|l|l|l|}\hline
$\kappa$ & $n_\mathrm{traj}$ & $n_\mathrm{step}$ & $dt$ & $\beta$ & $\gamma$ & $\gamma_b$ & $V_0$ & $d_0$  & $t_\mathrm{eq}$\\\hline
$0.1$ & $1.2 \cdot 10^4$ & $10^6$ & $10^{-2.5}$ & $1$ & $1$ & $10$ & $1$ & $1$ & $300$\\\hline
$1$ & $1.2 \cdot 10^4$ & $10^6$ & $10^{-2.7}$ & $1$ & $1$ & $10$ & $1$ & $1$ & $100$\\\hline
$10$ & $1.2 \cdot 10^4$ & $10^6$ & $10^{-2.9}$ & $1$ & $1$ & $10$ & $1$ & $1$ & $60$\\\hline
$100$ & $1.2 \cdot 10^4$ & $10^6$ & $10^{-3.4}$ & $1$ & $1$ & $10$ & $1$ & $1$ & $40$\\\hline
$1000$ & $1.2 \cdot 10^4$ & $10^6$ & $10^{-4}$ & $1$ & $1$ & $10$ & $1$ & $1$ & $20$\\\hline
\end{tabular}
\caption{\label{tab:mem_sim_dw}Simulation parameters for the double-well interaction potential.}
\end{ruledtabular}
\end{table}

\subsection{Stochastic Prandtl-Tomlinson model}

\begin{table}[H]
\begin{ruledtabular}
\begin{tabular}{|l|l|l|l|l|l|l|l|l|l|}\hline
$\kappa$ & $n_\mathrm{traj}$ & $n_\mathrm{step}$ & $dt$ & $\beta$ & $\gamma$ & $\gamma_b$ & $V_0$ & $d_0$  & $t_\mathrm{eq}$\\\hline
$0.1$ & $2.4 \cdot 10^3$ & $10^6$ & $10^{-2.5}$ & $1$ & $1$ & $10$ & $1$ & $1$ & $100$\\\hline
$1$ & $2.4 \cdot 10^3$ & $10^6$ & $10^{-2.7}$ & $1$ & $1$ & $10$ & $1$ & $1$ & $30$\\\hline
$10$ & $2.4\cdot 10^3$ & $10^6$ & $10^{-2.9}$ & $1$ & $1$ & $10$ & $1$ & $1$ & $10$\\\hline
$100$ & $2.4 \cdot 10^3$ & $10^6$ & $10^{-3.4}$ & $1$ & $1$ & $10$ & $1$ & $1$ & $5$\\\hline
$1000$ & $2.4 \cdot 10^3$ & $10^6$ & $10^{-3.6}$ & $1$ & $1$ & $10$ & $1$ & $1$ & $4$\\\hline
\end{tabular}
\caption{\label{tab:mem_sim_pt}Simulation parameters for the stochastic PT model.}
\end{ruledtabular}
\end{table}

\section{\label{app:FDT}Derivation of Eq.~\eqref{eq:memkernel_LT} and the fluctuation-dissipation theorem}
In this appendix, we derive Eq.~\eqref{eq:memkernel_LT} and prove the fluctuation-dissipation theorem by direct calculation for the overdamped generalized Langevin equation in Eq.~\eqref{eq:GLE_overdamped}. The route presented here is similar to the one shown in Ref.~\cite{shin2010brownian} for the underdamped case, but with the presence of an external harmonic field. In Eq.~\eqref{eq:GLE_overdamped} the initial preparation of the system has been shifted to the infinite past as is reflected by the lower integration boundary in the memory integral. 
In order to directly apply the convolution theorem to Eq.\ \eqref{eq:GLE_overdamped}, the equation of motion needs to be slightly recast according to
\begin{equation}
\int_0^t \dd{s} \Gamma(t-s)\dot{x}(s)=-\kappa x(t)+\Delta f(t)+f(t)\,,
\end{equation}
where we introduced a shift of the random force $\Delta f(t)=-\int_0^\infty\dd s\Gamma(t+s)\dot{x}(-s)$. We then obtain the equation of motion for the position autocorrelation function $C_{xx}(t)$ by multiplying both sides of the equation with $x(0)$ and subsequently perform an equilibrium average
\begin{equation}\label{eq:corrt_EOM}
\int_0^t \dd{s}\Gamma(t-s)\dot{C}_{xx}(s)=-\kappa C_{xx}(t)+\langle \tilde{f}(t) x(0)\rangle_\mathrm{eq}\,,
\end{equation}
where we used the abbreviation $\tilde{f}(t)=\Delta f(t)+f(t)$. In order to reach consistency with the Mori equation \cite{mori1965transport} (where the system is prepared at $t=0$ in a certain initial configuration), we require the relation $\langle \tilde{f}(t)x(0)\rangle_\mathrm{eq}=0$, i.e.\
\begin{equation}\label{eq:fx_corr}
\langle f(t)x(0)\rangle_\mathrm{eq}=\int_0^\infty \dd{s}\Gamma(t+s)\dot{C}_{xx}(-s)\,.
\end{equation}
By Laplace transforming Eq.~\eqref{eq:corrt_EOM} together with this relation, we derive Eq.~\eqref{eq:memkernel_LT} where the bath is prepared in thermal equilibrium at $t=0$. Replacing the lower integration boundary in the memory integral of Eq.~\eqref{eq:memkernel_LT} by the initial preparation time $t=0$ is thus at the cost of specifying initial conditions for particle and bath as was amply discussed in Ref.~\cite{shin2010brownian}.

It can now be shown by direct calculation that Eq.~\eqref{eq:GLE_overdamped} together with the relation in Eq.~\eqref{eq:fx_corr} implies the correct form of the fluctuation-dissipation theorem
\begin{equation}\label{eq:FDT}
\langle f(t)f(0)\rangle_\mathrm{eq}=k_B T\Gamma(t)\,, \qquad t>0\,,
\end{equation}
and hence correctly describes the equilibrium properties of the Brownian particle. Note that a similar derivation of the FDT has been presented in Ref.~\cite{lisy2019generalized} for the underdamped case by transforming the Mori equation into Laplace domain. In our proof no transformation to Laplace domain is necessary. We multiply both sides of Eq.~\eqref{eq:GLE_overdamped} with $f(0)$ and perform a subsequent equilibrium average to obtain
\begin{equation}
\langle f(t)f(0)\rangle_\mathrm{eq}=\kappa\langle f(t)x(0)\rangle+\int_{-\infty}^0\dd{s}\Gamma(-s)\langle f(t)\dot{x}(s)\rangle\,.
\end{equation}
By application of the requirement for the fluctuating force-position correlator in Eq.~\eqref{eq:fx_corr}, we may write
\begin{align}
\begin{split}
\langle f(t)f(0)&\rangle_\mathrm{eq}=-\kappa\int_0^\infty\dd{s}\Gamma(t+s)\dot{C}_{xx}(s)\\
{}&+\int_{-\infty}^0\dd{s}\int_0^\infty\dd{s'}\Gamma(-s)\dot{\Gamma}(t-s+s')\dot{C}_{xx}(s')\,.
\end{split}
\end{align}
A partial integration of the first term yields
\begin{align}
\begin{split}
\langle f(t)f(0)&\rangle_\mathrm{eq}=\kappa\Gamma(t)C_{xx}(0)\\
{}&-\int_0^\infty\dd{s}\int_0^s\dd{s'}\dot{\Gamma}(t+s)\Gamma(s-s')\dot{C}_{xx}(s')\\
{}&+\int_0^\infty\dd{s}\int_0^\infty\dd{s'}\Gamma(s)\dot{\Gamma}(t+s+s')\dot{C}_{xx}(s')\,.
\end{split}
\end{align}
Now using the fact that $C_{xx}(0)=\frac{k_B T}{\kappa}$ in equilibrium and realizing with some algebra that the last two terms compensate each other, we obtain the form of the FDT in Eq.~\eqref{eq:FDT}.

\section{\label{app:lin_response_v}Linear-response theory of a moving harmonic trap}
We want to derive the linear-response relation for the time-dependent friction coefficient $\gamma(t)$ as defined in Eq.~\eqref{eq:lin_response_v}. We consider a system of a harmonically trapped tracer particle that is arbitrarily coupled to a heat bath. The position of the minimum of the trap follows a time protocol $x_0(t)$ and we ask for its effect on the mean position of the tracer particle. The time-dependent external potential then reads
\begin{equation}
V_\mathrm{ext}=\frac{1}{2}\kappa(x-x_0(t))^2\,.
\end{equation}
For small dragging velocities $x_0(t)=\int_0^t\dd{t'} v_0(t')$ we may linearize the perturbation Hamiltonian
\begin{align}
H_\mathrm{pert}(t)=-\kappa x v_0t\theta(t)+\mathcal{O}(v_0^2)\,,
\end{align}
where we assumed the perturbation to be switched on at $t=0$ and the velocity to be constant over time. Direct application of the general form of the FDT yields \cite{hansen2013theory}
\begin{equation}
\langle x\rangle (t)=-\beta\int_0^t \dd{t'}\dv{}{t}\langle x(0)x(t-t')\rangle_\mathrm{eq}\kappa v_0 t'\,.
\end{equation}
By means of partial integration and the fact that $\langle x^2\rangle_\mathrm{eq}=\frac{k_B T}{\kappa}$ in equilibrium we then obtain
\begin{equation}
\langle x\rangle (t)-v_0 t=-\beta \kappa v_0\int_0^t \dd{t'} \langle x(0)x(t')\rangle_\mathrm{eq}\,.
\end{equation}
From this relation Eq.~\eqref{eq:lin_response_v} follows immediately.
\end{appendix}

\end{document}